\documentclass[aps,prb,twocolumn,showpacs,superscriptaddress,groupedaddress]{revtex4}
\usepackage{graphicx}  
\usepackage{dcolumn}   
\usepackage{bm}    
\usepackage{amssymb}

\begin{document}

\title{Feedback Control of Rabi Oscillations in  Circuit QED}

\author{Wei Cui}
\address{CEMS, RIKEN, Saitama, 351-0198, Japan}

\author{Franco Nori}
\address{CEMS, RIKEN, Saitama, 351-0198, Japan}
\address{Physics Department, The University of Michigan, Ann Arbor, Michigan 48109-1040, USA}

\date{\today}

\begin{abstract}
We consider the feedback stabilization of Rabi oscillations in a superconducting qubit which is coupled to a microwave readout cavity. The signal is readout by homodyne detection of the in-phase quadrature amplitude of the weak measurement output. 
By multiplying the time-delayed Rabi reference, one can extract the signal,  with maximum signal-to-noise ratio, from the noise. We further track and stabilize the Rabi oscillations by using Lyapunov feedback control to properly adjust the input Rabi drives.  Theoretical and  simulation results illustrate the effectiveness of the proposed control law.  
\end{abstract}
\maketitle


\section{Introduction}
In control theory, the system to be controlled is compared to the desired reference, and the discrepancy is used to correct the control action \cite{John1993}. In contrast to classical systems, where measurements do not alter the state of the system, quantum measurements will collapse the system instaneously into one of its eigenstates in a probabilistic manner: the ``measurement-induced backaction'' \cite{Schoelkopf20102}. Although the quantum coherent feedback control has been proposed \cite{James2009} and extensively applied in quantum optics and cooling mechanical oscillators and so on\cite{Mabuchi2012,Zhang2012,Xue2012}, the measurement-based feedback control still maintains a great interest.  Based on the quantum trajectory theory, Wiseman and Milburn  Ref.~\onlinecite{Wiseman2009} developed a quantum conditional stochastic master equation (SME)  to describe the dynamics resulting  from the feedback (of the measurement output at each instant) to the quantum system. SME has been a topic of considerable
activity in recent years for it paves the way for studying real-time measurement-based feedback control \cite{Wiseman2012,Haroche2011,Dicarlo2012,Qi2012} in quantum information processing and computation. 

Circuit quantum electrodynamics ({i.e.}, circuit QED, where a superconducting qubit is coupled to a microwave-frequency resonator cavity; see, e.g., Ref.~\onlinecite{You2005,You2011,Girvin2007,Ashhab2011}) has been shown to
be a promising quantum computing architecture. Circuit QED allows for rapid, repeated quantum nondemolition (QND) superconducting qubit measurement  \cite{Schoelkopf20102, Devoret2013} and also provides several simple high-fidelity readout mechanisms, such as using large measurement drive powers \cite{Schoelkopf2010}, and using either quantum-limited \cite{Siddiqi20123} or nonlinear bifurcation amplifiers \cite{Esteve2009}. Moreover, circuit QED is an excellent test-bed for implementing quantum feedback control in either the qubits or the microwave resonator \cite{Esteve2011,Johansson2010,Milburn2005,Milburn2010,Wallraff2010,Johansson2012,Korotkov2005,Cui2012,Szigeti2012}. For example, a recent work \cite{Siddiqi2012} has been shown that quantum measurement-based feedback control can reduce dephasing and remarkably prolong the Rabi oscillations.
 
 Here, we analytically derive a simple and experimentally-feasible measurement-based feedback control law for circuit QED to track and stabilize Rabi oscillations.  The paper is organized as follows. The next section contains a brief discussion of the circuit QED Hamiltonian, the quantum detection, and the stochastic master equation for the qubit. In Sec. III, we study the open-loop control of the Rabi oscillation. In Se. IV, we study the feedback control by the Lyapunov function method. We summarize our conclusions in Sec. V. 


 \begin{figure}
\centerline{\scalebox{0.48}[0.50]{\includegraphics{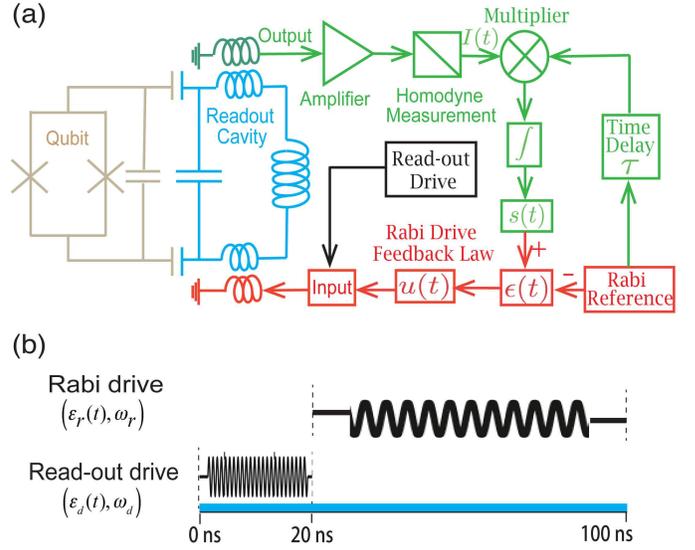}}}
\caption{(Color online) (a) Simplified circuit diagram of measurement and feedback control. A superconducting qubit (yellow) is coupled to a microwave readout cavity (blue). The amplified output is homodyne-detected and  the quadrature signal is then extracted from the noise by multiplying the time-delayed Rabi reference (green). The discrepancy is used to design the feedback control law to correct Rabi oscillations (red). (b) Schematic of the read-out drive to build up the photon population of the cavity and Rabi drive to stabilize the Rabi oscillation.}
\end{figure}

\section{circuit for measurement and feedback control}
As shown in Fig.1(a), we consider a superconducting circuit QED system with a superconducting qubit coupled to a microwave readout cavity and driven by two external drives: (i) a read-out drive with amplitude $\epsilon_{d}(t)$ and frequency $\omega_{d}$ near the cavity resonance frequency $\omega_{c}$, and (ii) a Rabi drive with amplitude $\epsilon_{r}(t)$ and frequency $\omega_{r}$ near the frequency of the qubit $\omega_{q}$,
\cite{Siddiqi2012,Esteve2009,Oliver2005,Martinis2013}. The Hamiltonian of the entire system can be written as 
\begin{eqnarray}
H & = & \hbar\omega_{c}a^{\dagger}a+\hbar\frac{\omega_{q}}{2}\sigma_{z}+\hbar g(a^{\dagger}\sigma_{-}+a\sigma_{+})\nonumber \\
 &  & +\hbar[\epsilon_{d}(t)e^{-i\omega_{d}t}a^{\dagger}+\epsilon_{r}(t)e^{-i\omega_{r}t}a^{\dagger}+\mathrm{h.c}],\label{Hamiltonian}
\end{eqnarray}
where $a^{\dagger}$ and $a$ are the creation and annihilation operators
for the microwave readout cavity, $\sigma_{+}$ and $\sigma_{-}$ are the raising
and lowering operators of the superconducting qubit, and $g$ is the
coupling strength between the cavity and the qubit. In the dispersive
regime \cite{Siddiqi20122}, $\left|\Delta\right|=\left|\omega_{q}-\omega_{c}\right|\gg g$,
by applying the dispersive shift $U=\exp[{g(a\sigma_{+}-a^{\dagger}\sigma_{-})/\Delta}]$,
and moving to the rotating frames for both the qubit and cavity, $U_{c}=\exp[{-ia^{\dagger}a\omega_{d}t}]$,
$U_{q}=\exp[{-i\sigma_{z}\omega_{r}t/2}]$, with the
rotating-wave approximation, the Hamiltonian in Eq. (\ref{Hamiltonian})
becomes 
\begin{eqnarray}
H_{\mathrm{eff}} & = & \hbar\Delta_{c}a^{\dagger}a+\hbar\chi a^{\dagger}a\sigma_{z}+\hbar\frac{\tilde{\omega}_{q}}{2}\sigma_{z}+\hbar\frac{\Omega_{R}}{2}\sigma_{x}\nonumber \\
 &  & +\hbar\left[\epsilon_{d}(t)a^{\dagger}+\epsilon_{d}^{\ast}(t)a\right],\label{Effective Hamiltinian}
\end{eqnarray}
where 
$\Delta_{c}=\omega_{c}-\omega_{d}, \chi=g^{2}/\Delta, \Omega_{R}=2\epsilon_{r}(t)g/\Delta$
and the Lamb-shifted qubit transition frequency $\tilde{\omega}_{q}=\omega_{q}-\omega_{r}+\chi$.
%
%
%

If the cavity state is coherent, and the microwave cavity decay rate is much larger than the qubit decay rate, $\kappa\gg\gamma_{1}$ (that
allows to decouple the qubit dynamics from the resonator adiabatically),
the state at time $t$ is given by $\left|g\right\rangle \otimes\left|\alpha_{g}(t)\right\rangle $
or $\left|e\right\rangle \otimes\left|\alpha_{e}(t)\right\rangle $.
Here $\left|\alpha_{g(e)}(t)\right\rangle $ are coherent states
of the cavity and, from Eq.~(\ref{Effective Hamiltinian}),
the field amplitudes are given by \cite{Gambetta2008},
\begin{eqnarray}
\dot{\alpha}_{g}(t) & = & -i\epsilon_{d}(t)-i(\Delta_{c}-\chi)\alpha_{g}(t)-\frac{\kappa}{2}\alpha_{g}(t),\nonumber \\
\dot{\alpha}_{e}(t) & = & -i\epsilon_{d}(t)-i(\Delta_{c}+\chi)\alpha_{e}(t)-\frac{\kappa}{2}\alpha_{e}(t).\label{resonator equation}
\end{eqnarray}
Thus, these coherent states $\alpha_{g(e)}$ act as ``pointer states" \cite{Wiseman2009} for the qubit. 
Based on homodyne detection, by applying the transformation 
$$P(t)=|e\rangle\langle e|D[\alpha_e(t)]+|g\rangle\langle g|D[\alpha_g(t)],$$
 with $D[\alpha]=\exp[\alpha a^{\dag}-\alpha^{\ast}a]$ as the displacement operator of the microwave cavity, the effective stochastic master equation for the qubit degrees of freedom is 
\begin{eqnarray}
d\tilde{\rho}&=&-\frac{i}{\hbar}\frac{\tilde{\omega}_{ac}(t)}{2}\left[\sigma_{z},\tilde{\rho}\right]dt-i\frac{\Omega_{R}}{2}\left[\sigma_{x},\tilde{\rho}\right]dt+\gamma_{1}\mathcal{D}\left[\sigma_{-}\right]\tilde{\rho}dt\nonumber \\
&+&\frac{\gamma_{\phi}+\Gamma_{d}(t)}{2}\mathcal{D}\left[\sigma_{z}\right]\tilde{\rho}dt+\sqrt{\kappa\eta}\left|\beta(t)\right|\mathcal{H}\left[\sigma_{z}\right]\tilde{\rho}dW_{t}.\label{SME}
\end{eqnarray}
Here $$\tilde{\omega}_{ac}(t)=\tilde{\omega}_{q}+B(t),$$ 
and
 $$\beta(t)=\alpha_{e}(t)-\alpha_{g}(t)$$
is the separation between the pointer states $\alpha_g(t)$ and $\alpha_e(t)$, $\eta$ is the measurement
efficiency, $\gamma_{\phi}$ is the pure dephasing rate,  $\mathcal{D}[A]$
is the damping superoperator 
$$\mathcal{D}[A]\rho=A\rho A^{\dagger}-A^{\dagger}A\rho/2-\rho A^{\dagger}A/2,$$ 
 and $$\mathcal{H}\left[A\right]\tilde{\rho}=A\tilde{\rho}+\tilde{\rho}A^{\dagger}-\left\langle A+A^{\dagger}\right\rangle \tilde{\rho}.$$
Also, $$\Gamma_{d}(t)=2\chi\mathrm{Im}[\alpha_{g}(t)\alpha_{e}^{\ast}(t)]$$
is the measurement-induced dephasing and $$B(t)=2\chi\mathrm{Re}[\alpha_{g}(t)\alpha_{e}^{\ast}(t)]$$
is the ac Stark shift. The innovation $dW_{t}$
is a Wiener process \cite{Wiseman2009} with $$\mathrm{E\left[\mathit{dW_{t}}\right]}=0,~
\text{and} ~\mathrm{E}[dW_{t}^{2}]=dt.$$ Due to the qubit decay $\gamma_{1}$ and dephasing $\gamma_{\phi}+\Gamma_{d}(t)$, the system must quickly lose its quantum features.

A coherent drive is turned on for 20 ns to build up the photon population of the cavity and is then repeated every 100 ns (see Fig.1(b)).
 The cavity pull is designed to be $\chi/2\pi=5$ MHz, and the cavity decay rate is $\kappa/2\pi=20$ MHz. 
A homodyne detection of the readout cavity field, with the help of the distance $\beta(t)$ between the states $|\alpha_{e}(t)\rangle$ and $|\alpha_{g}(t)\rangle$, can then be used to distinguish the coherent states and thus readout the state of the qubit. 
By applying the $P$-transformation to the in-phase quadrature amplitude $$I_{\phi}=\left\langle ae^{-i\phi}+a^{\dagger}e^{i\phi}\right\rangle /2,$$ with $\phi$ the phase of the local oscillation, 
 the
%
homodyne measurement record coming from the microwave cavity becomes 
\begin{equation}
I(t)=\sqrt{\kappa\eta}\left|\beta(t)\right|\left\langle \sigma_{z}(t)\right\rangle+\xi(t)=s(t)+\xi(t),\label{current}
\end{equation}
 where the qubit uncorrelated term $\sqrt{\kappa\eta}|\mu(t)|\sin[\phi+\arctan(\mu)]$, $\mu=\alpha_g+\alpha_e$, has been omitted. We have set the homodyne phase $\phi$ to $\arg(\beta)$, which corresponds to detecting the quadrature with the greatest separation of the pointer states. Here $\xi(t)=dW_t/dt$ is a Gaussian white noise, representing the shot noise, with spectral density $P_{\xi}(\omega)=1$. Usually, the quantum signal $s(t)$ is very weak and the noise  $\xi(t)$  may be strong. The overall objective is to make the system behave in a desired way by manipulating the input drive based on the measurement output.
 Here we expect to sustain the Rabi oscillations. To achieve this, the following steps are required: Detect the signal $s(t)$ from the noise $\xi(t)$; reconstruct $x_{1}(t)=\mathrm{Tr}[\sigma_{x}\tilde{\rho}(t)]$, $x_{2}(t)=\mathrm{Tr}[\sigma_{y}\tilde{\rho}(t)]$,
and $x_{3}(t)=\mathrm{Tr}[\sigma_{z}\tilde{\rho}(t)]$, which are the three components of the Bloch vector for the ensemble qubit state based on the detected signal \cite{Liu2005}; feedback the error signal between reconstructed state and the desired state, to design the feedback control law (the Rabi drive) thus minimizing the error.

\begin{figure}
\centerline{\scalebox{0.5}[0.50]{\includegraphics{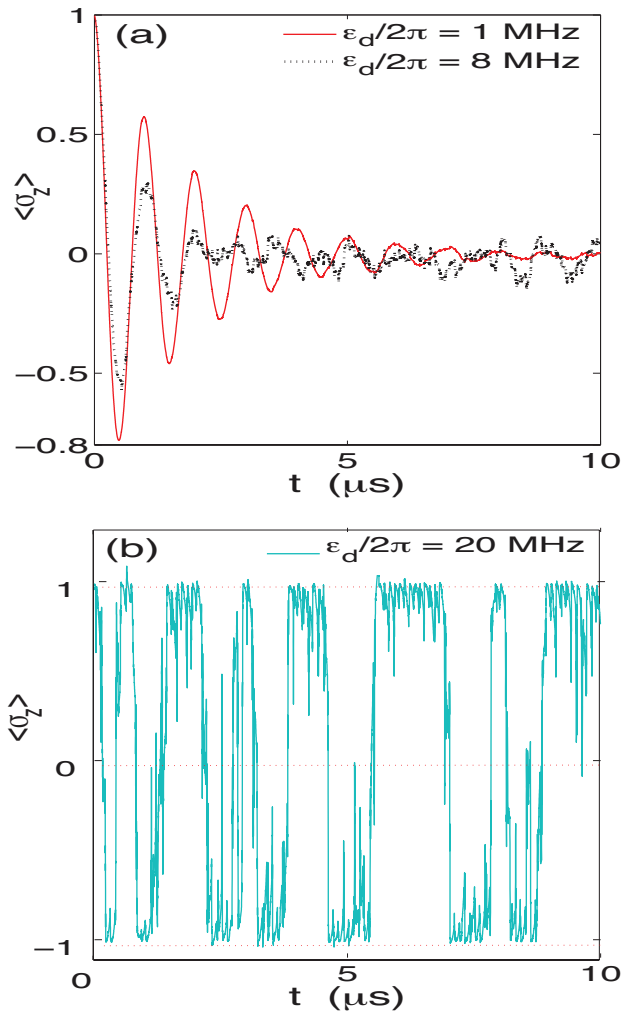}}}
\caption{(Color online) (a)  ensemble-averaged behavior (over 1,000 realizations) of  the filtered signal (only plot $\langle\sigma_z(t)\rangle$) by continuous weak measurements of the open-loop controlled microwave readout cavity with Rabi frequency $\Omega_R/2\pi=1$ MHz. The Rabi-drive amplitude $\epsilon_r=\Omega_R\delta/2g$ and the frequency $\omega_r=\omega_q+\chi$, which makes the Lamb-shifted qubit transition frequency equals zero. The qubit is initially in the excited state and the read-out drive amplitude are $\epsilon_d/2\pi=1$ MHz (red solid line), and 8 MHz (black dotted line), respectively. (b) shows the same quantity except just a single realization by strong measurement with measurement drive amplitude of amplitude $\epsilon_d/2\pi=20$ MHz. }
\end{figure}


\section{Open--loop control: no feedback}
To see how the feedback Rabi drive will work, we first consider the open-loop control.  Open-loop means that we do not use feedback to determine if the output has achieved the desired goal.  One can simply drive the microwave cavity with amplitude $$\epsilon_r=\Omega_R\delta/2g$$ to obtain the Rabi oscillation with Rabi frequency $\Omega_R$; but cannot correct any errors.
 To illustrate this, we have numerically simulated the microwave cavity field equation (3) and the superconducting qubit  stochastic master equation (4) with the open-loop drive amplitude $\epsilon_r=\Omega_R\delta/2g$ to obtain the expected Rabi frequency $\Omega_R/2\pi=1$ MHz, for four different measurement drives $\epsilon_d/2\pi=1$ MHz, 4 MHz,  8 MHz, and 20 MHz. 

In Fig. 2, we show some of these numerical results for the open-loop control of Rabi oscillations with frequency $\Omega_R/2\pi=1$ MHz. We set the initial state of the qubit as the excited state. 
Figure 2(a) shows the results averaged over 1000 realizations.  In these results we set the measurement efficiency $\eta=1$, the qubit decay   $\gamma_1/2\pi=0.05 $ MHz, and the pure dephasing rate  $\gamma_{\phi}/2\pi=0.1$ MHz. 
The Rabi-drive amplitude $\epsilon_r=\Omega_R\delta/2g$ and the frequency $\omega_r=\omega_q+\chi$,  should be chosen carefully to make the Lamb-shifted qubit transition frequency equal zero.  When acquiring information from the measurement, it of course induces significant backaction on the system. From Fig.~2(a), we see that for the small measurement-drive amplitude  ($\epsilon_d/2\pi=1$ MHz, red solid curve), the qubit decays and pure dephasing  dominates the evolution. Thus, in this case, the measurement only causes small amplitude noise on the Rabi oscillation. However, for the larger drive amplitude $\epsilon_d/2\pi=4$ MHz (not shown) and 8 MHz (black dotted curve) the measurement induces remarkable backaction on the qubit.  

We set $\epsilon_d/2\pi=20$ MHz to gain more insight into what is actually happening during the evolution of the Rabi oscillation with strong measurement-drive amplitude. As shown in Fig.~2, $\langle\sigma_z\rangle$ exhibits decaying oscillations, in Fig. 2(a), when the drive is weak ($\epsilon_d/2\pi=1$,~and 8 MHz) and discontinuous jumps between two levels, in Fig. 2(b), when the driving is strong ($\epsilon_d/2\pi=20$ MHz).
Clearly, in the strong drive, the qubit will remain fixed in either $z=+1$ or $-1$. This is the Zeno effect. All these demonstrated that the open-loop control cannot compensate for the disturbances in the system.  

\section{Feedback control}
We now propose a simple feedback control law allowing to compensate the dephasing of the superconducting qubit, the measurement-induced backaction, and  to maintain the coherence of the Rabi oscillations based on the above measurement scheme. The schematics of such feedback control is shown in Fig.~1. The amplified and filtered signal $s(t)=\langle\sigma_z(t)\rangle$ is compared with the Rabi reference signal $s^{\ast}(t)=\cos\Omega_R^0t$, and the difference 
\begin{eqnarray}\label{error}
\varepsilon(t)=s(t)-s^{\ast}(t)
\end{eqnarray}
is used to generate the feedback signal $u(t)$ that drives the microwave cavity in order to reduce the difference with the desired Rabi oscillations: $\varepsilon(t)\to0$ (frequency tracking \cite{Kurt2011}). The difference $\varepsilon(t)$ evolves as
\begin{eqnarray}\label{derror}
\dot{\varepsilon}(t)&=&\dot{s}(t)-\dot{s}^{\ast}(t)=\mathop{\rm E}\left[\frac{d}{dt}\langle\sigma_z(t)\rangle\right]-\dot{s}^{\ast}(t)\\
&=&\Omega_R(t)\langle\sigma_y(t)\rangle-\gamma_1(1+\langle\sigma_z(t)\rangle)+\Omega_R^0\sin\Omega_R^0t.\nonumber
\end{eqnarray}
Thus, we design the feedback control law (the Rabi-drive amplitude):
\begin{eqnarray}\label{control}
u(t)=&\epsilon(t)&=-\frac{\delta}{2g}\langle\sigma_y(t)\rangle^{-1}\left[K_1 \mathop{\rm sign}\varepsilon(t)+K_2\varepsilon(t)\right.\nonumber\\
&&\left.-\gamma_1(1+\langle\sigma_z(t)\rangle)+\Omega_R^0\sin\Omega_R^0t\right],
\end{eqnarray}
where $K_1, K_2>0$. Using the feedback-control law  (\ref{control}) in Eq. (\ref{derror}), we have
\begin{eqnarray}
\dot{\varepsilon}(t)=-K_1 \mathop{\rm sign}\varepsilon(t)-K_2\varepsilon(t).
\end{eqnarray}
Clearly, if $\varepsilon(t)>0$, then $\dot{\varepsilon}(t)<0$; and if $\varepsilon(t)<0$, then $\dot{\varepsilon}(t)>0$.

The Lyapunov function method \cite{Mirrahimi2005,Dong2012} is usually employed to prove the stability of an ordinary differential equation and widely used in stability and control theory. Here we can choose a simple Lyapunov function $$\nu(t)=\varepsilon^2(t)/2.$$ Obviously, $$\nu(t)>0 ~\text{and}~ \dot{\nu}(t)=\dot{\varepsilon}(t)\varepsilon(t)<0.$$ 
Then, the Lyapunov theorem tells us that every trajectory of Eq. (\ref{error}) converges to zero:
\begin{eqnarray}
\lim_{t\to\infty}|s(t)-s^{\ast}(t)|\to0~~~~{\rm as} ~~~~t\to\infty,
\end{eqnarray}
which means the system is globally asymptotically stable. Now, the only problem is to choose $K_1$ and $K_2$. From the feedback control law in Eq. (\ref{control}), we find that when $s(t)$ is far from $s^{\ast}(t)$,  a large $K_2$ is needed to make $s(t)$ converge to $s^{\ast}(t)$ quickly.  If $s(t)$ is quite close to $s^{\ast}(t)$, $\mathop{\rm sign}\varepsilon(t)$ dominates the evolution, thus a small $K_1$ is needed to reduce the error $\varepsilon(t)$. 

\begin{figure}
\centerline{\scalebox{0.38}[0.45]{\includegraphics{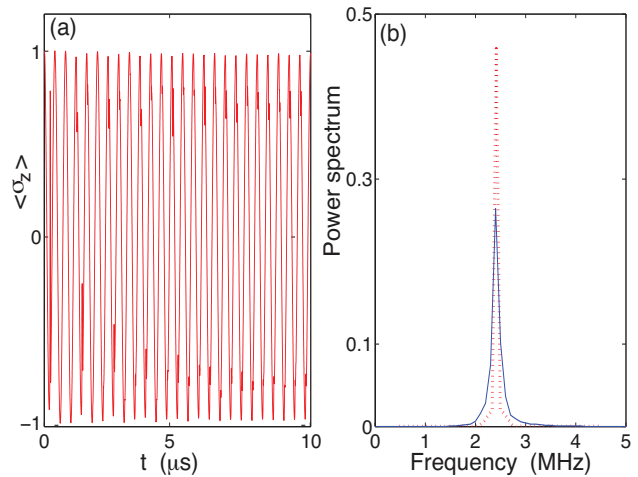}}}
\caption{(Color online) (a) Feedback-controlled ensemble-averaged (over 1,000 realizations) Rabi oscillations, which persist for much longer time than those with open-loop control. The Rabi frequency $\Omega_R^0/2\pi=2.5$ MHz and the read-out drive amplitude is $\epsilon_d/2\pi=1$ MHz.  (b) Power spectral density for the averaged measurement of feedback-controlled Rabi oscillations from (a) (red curve); the blue curve corresponds to the open-loop case with the same parameters of (a).   }
\end{figure}

We have simulated the feedback loop designed above to maintain the Rabi oscillations with frequency 
$\Omega_R^0/2\pi=2.5$ MHz. The measurement is set in the weak-driving regime, when the readout drive amplitude is $\epsilon_d/2\pi=1$ MHz, where the measurement-induced backaction $\Gamma_d(t)$ and $B(t)$ remain small. The control parameters $K_1=5\times10^6$ and $K_2=10^8$. The other parameters are the same as in the case of open-loop control. Figure 3(a) shows typical realizations of the feedback-controlled ensemble-averaged Rabi oscillations. Clearly, the feedback control can quickly track the reference Rabi signal and ideally fight against dephasing and the measurement-induced backaction. From Fig.~3 we can see that the feedback-controlled Rabi oscillations persist for much longer time than those with open-loop control. Finally, in Fig.~3(b), we  compare the power spectral density of the averaged measurement record in feedback-controlled Rabi oscillations (red curve) with the corresponding open-loop control (blue curve).  Both of them are centered at 2.5 MHz. However, the feedback controlled spectrum has a needle-like peak at the Rabi reference frequency, while the open-loop controlled spectrum has a broad distribution. Thus, we can precisely convert the amplitude of the Rabi microwave drive to a frequency. 
Clearly, the proposed feedback control has more advantages than the open-loop control, for stabilizing the Rabi oscillations in circuit QED.

\section{conclusion}
In conclusion, we have proposed and analyzed a quantum feedback control method to stabilize the Rabi oscillations in a superconducting qubit which is coupled to a microwave readout cavity. The control law can be conveniently tested in realistic quantum QED architectures. The output signal detection has been discussed and the maximum signal-to-noise ratio has been given. We have also analytically proven that the designed feedback Rabi-drive amplitude can make the averaged filtered signal  quickly converge to the reference Rabi signal. We have discussed the advantages of the quantum feedback control, over the open-loop control, in stabilizing the Rabi oscillations. The proposed Lyapunov feedback control can be further applied to quantum state purification, quantum adaptive measurement, and quantum parameter estimation.

%
%
%
%
%

\begin{acknowledgements}
WC is supported by the RIKEN FPR Program. FN is partially supported
by the ARO, JSPS-RFBR Contract No. 12-02-92100, a Grant-in-Aid for
Scientific Research (S), MEXT ``Kakenhi on Quantum Cybernetics'',
and the JSPS via its FIRST program.
\end{acknowledgements}

\label{sec:TeXbooks}

\end{document}